\documentclass{ws-ijmpa}

\begin{document}

\markboth{Gustavo Dotti, Julio Oliva and Ricardo Troncoso}
{Vacuum solutions with nontrivial boundaries for Einstein-Gauss-Bonnet}

%
\catchline{}{}{}{}{}
%

\title{Vacuum solutions with nontrivial boundaries\\ for the Einstein-Gauss-Bonnet theory}

\author{Gustavo Dotti}

\address{Facultad de Matem\'{a}tica, Astronom\'{\i}a y F\'{\i}sica, Universidad
Nacional de C\'{o}rdoba,\\ Ciudad Universitaria, (5000) C\'{o}rdoba, Argentina.}

\author{Julio Oliva}

\address{Centro de Estudios Cient\'{\i}ficos (CECS), Casilla 1469, Valdivia, Chile.}

\author{Ricardo Troncoso}

\address{Centro de Estudios Cient\'{\i}ficos (CECS), Casilla 1469, Valdivia, Chile.\\
Centro de Ingenier\'{\i}a de la Innovaci\'{o}n del CECS (CIN), Valdivia, Chile.}

\maketitle

\begin{abstract}
The classification of certain class of static solutions for the
Einstein-Gauss-Bonnet theory in vacuum is presented. The spacelike section of the class of metrics under
consideration is a warped product of the real line with a nontrivial base
manifold. For arbitrary values of the Gauss-Bonnet coupling, the base manifold must be Einstein with an
additional scalar restriction. The geometry of the boundary can be relaxed only when the Gauss-Bonnet coupling is related with the cosmological and Newton constants, so that the theory admits a unique maximally symmetric solution. This additional freedom in the boundary metric allows the existence of three main branches of geometries in the bulk, containing new black holes and wormholes in vacuum.

\end{abstract}

\ccode{PACS numbers: 04.50.+h, 04.20.Jb, 04.90.+e}

The basic principles of General Relativity in dimensions higher than four give rise
to the Lovelock theories of gravity \cite{Lovelock}, with the most general covariant, divergency free second order
field equations for the metric. In three and four dimensions, the Lovelock theory
reduces to General Relativity with cosmological constant, while in five and six dimensions it corresponds to
the so-called Einstein-Gauss-Bonnet \textbf{(EGB)} theory, whose field equations contain quadratic
powers of the curvature in a very precise combination. In $d$ dimensions, the field equations of the EGB theory read
\vspace{-.15cm}
\begin{gather}
\mathcal{E}_{a}:=\epsilon_{ab_{1}...b_{d-1}}\!\left[  \left(  d-4\right)
\alpha_{2}R^{b_{1}b_{2}}R^{b_{3}b_{4}}\!+\left(  d-2\right)  \alpha
_{1}R^{b_{1}b_{2}}e^{b_{3}}e^{b_{4}}\!\right.  \nonumber\\
\ \ \ \ \ \ \ \ \ \ \ \ \ \ \ \ \ \ \ \ \ \ \ \ \ \ \ \ \ \ \ \ \ \ \ \ \ \ \ \ \ \ \ \ \ \ \ \ \ \ \ \ \left.
+d\ \alpha_{0}e^{b_{1}}e^{b_{2}}e^{b_{3}}e^{b_{4}}\right]e^{b_{5}}\!...e^{b_{d-1}%
}  =0\label{feq}%
\end{gather}
where $e^{a}$ is the vielbein and $R^{ab}$ stands for the curvature two-form.
The last two terms in (\ref{feq}) correspond to the Einstein tensor and the cosmological term, respectively, so that General Relativity is recovered when $\alpha_{2}=0$. This theory possesses a wide spectrum of solutions in vacuum\cite{BD,CaienAdS} including static wormholes\cite{DOTWorm}$^{-}$\cite{CamelloMarameo2}, black holes with nontrivial boundaries\cite{DOT5d,Cai,BHStopo}, black p-branes\cite{GOT,KM}, solutions with NUT charge\cite{Mann}, nontrivial torsion\cite{ArosContreras}$^{-}$\cite{Camello2}, spontaneous compactifications\cite{MH}$^{-}$\cite{ArosRomo} and
metrics with a nontrivial jump in the extrinsic curvature\cite{HTronZ,GGGW}.

 For generic values of the couplings, the theory admits two maximally symmetric solutions, which merge to a single one provided the Gauss-Bonnet coupling in Eq.(\ref{feq}) fulfils\cite{BHS}
\begin{equation}
\alpha_{2}=\frac{\left(  d-2\right)  ^{2}}{4d\left(  d-4\right)  }\frac
{\alpha_{1}^{2}}{\alpha_{0}}.\label{tuneo}%
\end{equation}
In Ref.\refcite{DOT5d}, vacuum solutions for the five-dimensional EGB theory were classified within the following class of metrics:
\begin{equation}
ds^{2}=-f^{2}\left(  r\right)  dt^{2}+\frac{dr^{2}}{g^{2}\left(  r\right)
}+r^{2}d\Sigma_{d-2}^{2} ,\label{metric}%
\end{equation}
where $d\Sigma_{d-2}$ is the line element of an arbitrary $(d-2)$-dimensional base manifold which determines the geometry at the boundary. Here we extend the analysis of Ref.\refcite{DOT5d} to dimensions greater than five. For simplicity we focus on the six-dimensional case, which captures the main features of the entire classification. The extension of these results to higher dimensions is performed in detail in Ref.\refcite{DOTHd}.

\section{Classification of the six-dimensional case}
\subsection{Generic case}
For arbitrary values of the coupling constants, the most general solution in vacuum for the EGB equation (\ref{feq}) in six dimensions within the family of metrics given by (\ref{metric}) reduces to
\begin{equation}
f^{2}\left(  r\right)  =g^{2}\left(  r\right)  =\gamma+\frac{\alpha_{1}%
}{\alpha_{2}}r^{2}\left[  1\pm\sqrt{\left(  1-3\frac{\alpha_{2}\alpha_{0}%
}{\alpha_{1}^{2}}\right)  +\frac{\mu}{r^{5}}+\frac{\alpha_{2}^{2}}{\alpha
_{1}^{2}}\frac{\left(  \gamma^{2}+\xi\right)  }{r^{4}}}\right]  \ ,\label{generic}
\end{equation}
where the base manifold $\Sigma_{4}$ must be Einstein, i.e. $\tilde{R}_{~j}^{i}=3\gamma\delta_{j}^{i}$, with the following scalar condition:
\begin{equation}
\tilde{R}_{\ kl}^{ij}\tilde{R}_{~ij}^{kl}-4\tilde{R}_{ij}\tilde{R}^{ij}%
+\tilde{R}^{2}+24\xi=0,\label{E4propxi}%
\end{equation}
where $\tilde{R}_{\ kl}^{ij}$, and  $\tilde{R}_{ij}$ are the Riemann and Ricci tensors of $\Sigma_4$, respectively, and $\tilde{R}$ correspods to its Ricci scalar. This last condition means that the Euler density of $\Sigma_{4}$ must be constant. Thus, assuming the base manifold $\Sigma_{4}$ to be compact without boundary, integration of Eq.(\ref{E4propxi}) on $\Sigma_{4}$ gives a topological restriction on the base manifold, which reads $\xi=-\frac{4}{3}\pi^{2}\frac{\chi(\Sigma_{4})}{\mathcal{V}_{4}}$. Here $\chi(\Sigma_{4})$ is the Euler characteristic of the base manifold and $\mathcal{V}_{4}$ stands for its volume.
Note that the term proportional to $r^{-4}$ inside the square root of (\ref{generic}) does not appear in the Boulware-Deser solution\cite{BD,Cai} since it vanishes if and only if the base manifold is of constant curvature. It is worth pointing out that this term severely modifies the asymptotic behavior of the metric. Depending on the value of the parameters, this spacetime can describe black holes being asymptotically locally (A)dS or flat.
\subsection{Special case}
In six dimensions, in the special case in which the Gauss-Bonnet coupling is given by (\ref{tuneo}), the solutions split into four main branches according to the geometry of the base manifold:
\\
\textbf{First branch:} The base manifold $\Sigma_4$ has the same restrictions as in the generic case, i.e., it is Einstein and
satisfies (\ref{E4propxi}). The entire metric can describe black holes with nontrivial boundaries. The results of Ref.\refcite{BHStopo} are recovered if we further restrict to  base manifolds of constant curvature.\\

The special case, however,  allows the possibility of relaxing the condition that the base manifold be Einstein, as long as
\begin{equation} \label{forall}
g^2(r) =\sigma r^2+\gamma, \hspace{1cm} \sigma:=3\alpha_0/\alpha_1,
\end{equation}
\begin{equation}
\tilde{R}_{\ kl}^{ij}\tilde{R}_{~ij}^{kl}-4\tilde{R}_{ij}\tilde{R}^{ij}
+\tilde{R}^{2}-4\gamma\tilde{R}+24\gamma^{2}=0, \label{paratodas}
\end{equation}
case in which we get the three remaining branches:
\\
\textbf{Second branch (Special class of black holes):} If the we do not impose further requirements on $\Sigma_4$ besides  Eq.(\ref{paratodas}),
 the solution is given by Eqs. (\ref{metric}) and (\ref{forall}) with $f^2 = g^2$.
\\
\textbf{Third branch (Wormholes and spacetime horns):} For compact base manifolds without boundary satisfying (\ref{paratodas}) and
having a constant Ricci scalar, $\tilde{R}=12\gamma$, the integration of  (\ref{paratodas}) on $\Sigma_4$
assuming $\gamma=\pm 1$ gives  $\chi(\Sigma_{4})=\frac{3}{4\pi^{2}}\mathcal{V}_{4}$, where $\chi(\Sigma_{4})$ is the Euler characteristic of the base manifold,
 which, being proportional to its volume $\mathcal{V}_{4}$, cannot be negative. In this case the metric is given by Eq.(\ref{metric}) with $g^2(r)=\sigma r^2+\gamma$ and
\begin{equation}
f^{2}(r)=\left\{
\begin{array}
[c]{ccc}%
\left(  a\sqrt{\sigma r^{2}-1}+1-\sqrt{\sigma r^{2}-1}\tan^{-1}\left(
\frac{1}{\sqrt{\sigma r^{2}-1}}\right)  \right)  ^{2} & : & \gamma=-1\\
\left(  a\sqrt{\sigma r^{2}+1}+1-\sqrt{\sigma r^{2}+1}\tanh^{-1}\left(
\frac{1}{\sqrt{\sigma r^{2}+1}}\right)  \right)  ^{2} & : & \gamma=1
\end{array}
\right. ~,\label{fcuaworm}%
\end{equation}
where $a$ is an integration constant. It is
simple to show that, for negative cosmological constant $\left(
\sigma>0\right)  $ and $\gamma=-1$, the spacetime can be extended to describe a static wormhole solution
in vacuum \cite{DOTHd}, provided $a^{2}<\frac{\pi^{2}}{4}$.
In the case $\gamma=0$, $\tilde R =0$
 and, for compact $\Sigma_4$ without boundary, $\chi(\Sigma_{4})=0$. In this case
 the solution is given by (\ref{metric}), with $g^2(r)=\sigma r^2$ and $f(r)=\left(  a\sqrt{\sigma}r+\frac{1}{\sqrt{\sigma}r^{2}}\right)$.
If $\sigma>0$ and $a\geq0$ this
looks like a \textquotedblleft spacetime horn".
\\
\textbf{Fourth branch:} if  $\Sigma_4$ is a constant curvature manifold, the field equations degenerate and
$f$ can be an arbitrary function.\\

In sum, we have classified the solutions of the EGB theory in vacuum within the class of metrics (\ref{metric}). In the generic case, the base manifold must be Einstein with constant Euler density. This scalar condition is written in (\ref{E4propxi}), and it can be seen as the six-dimensional analogue of the restriction on the squared Weyl tensor found in Ref. \refcite{DG}. It was shown that the base manifold acquires additional freedom only for the special case (\ref{tuneo}), where new wormholes in vacuum and black holes with nontrivial boundaries are found. The extension of these results to higher dimensions involves, among further details, a new parameter that characterizes the geometry of the boundary\cite{DOTHd}.

\section*{Acknowledgments}

 This work was
partially funded by Secyt-UNC\'ordoba, CONICET (Argentina); FONDECYT grants 1051056, 1061291, 1071125, 1085322, 3085043. The Centro
de Estudios Cient\'{\i}ficos (CECS) is funded by the Chilean Government
through the Millennium Science Initiative and the Centers of Excellence Base
Financing Program of CONICYT. CECS is also supported by a group of private
companies which at present includes Antofagasta Minerals, Arauco, Empresas
CMPC, Indura, Naviera Ultragas and Telef\'{o}nica del Sur. CIN is funded by
CONICYT and the Gobierno Regional de Los R\'{\i}os. G.D. is partially supported by CONICET.

\end{document}